\documentstyle[12pt]{article}

\input master.ipt
\input masterxt.ipt

\def\numt{\mbox{$\nu_{\mu(\tau)}$}} 
                                 
\def\PTot{\mbox{$P_\oplus$}}

\def\AsymSs{\mbox{$A_S^s$}}

\def\Ses{\mbox{$\calS_e^s$}}
\def\Res{\mbox{$\calR_E^s$}}

\def\FigSKSpectra{1}   
\def\FigSKDist{2}      

\def\TabEventRates{I}  

\begin{document}
\sloppy


\vspace{0.5cm}
%
%
{\normalsize
\begin{flushright}
\begin{tabular}{l}
arch-iv/9703207
\end{tabular}
\end{flushright}
}
\vspace{1cm}

\begin{center}
{\Large
Predictions for Spectral Distortions, Event Rates and \\
\vspace{0.25cm}
D-N Asymmetries for the \SK\ Detector\\
}
\end{center}

\begin{center}
Michele Maris $^{\mbox{a,b)}}$,
Serguey T. Petcov 
\footnote{
Also at: Institute of Nuclear
Research and Nuclear Energy, Bulgarian Academy of Sciences,
1784 Sofia, Bulgaria.} $^{\mbox{a,c)}}$ \\
$^{\mbox{a)}}$ Scuola Internazionale Superiore di Studi Avanzati,
Trieste, Italy.\\
$^{\mbox{b)}}$ INFN - Sezione di Pavia, Pavia, Italy.\\
$^{\mbox{c)}}$ INFN - Sezione di Trieste, Trieste, Italy.
\end{center}


\bec
\abstract{
\noindent
Day - Night asymmetries and electronic spectra distortions for the \SK\
detector 
are computed for the transition $\nu_e \rightarrow \nu_{\mu(\tau)}$\
and for a set of neutrino parameters.
The results show the possibility to enhance up to a factor 
six the Day - Night asymmetry 
selecting neutrino events induced by neutrinos 
which cross the Earth core.
This should increases the sensitivity of the \SK\ detector to the
Day - Night effect for $\sin^2 2 \theta_v < 0.01$\ and 
$5 \times 10^{-6}$ eV$^2 \, \lsim \, \Delta m^2 \, \lsim \, 10^{-5}$ eV$^2$.
}
\eec

\section{Introduction}
In \cite{ArticleI}\ the probabilities for the MSW conversion in Sun and Earth
for solar neutrinos $\PTot(\nue\rightarrow\numt)$\ were computed for the \SK\ 
experiment with a $1\%$\ accuracy.
Probabilities were computed separately for the day, the night and for the 
sample of neutrino events which are 
induced by neutrinos whose trajectories cross the Earth core.
In this way the probability for the Earth effect is enhanced up to a factor six 
for $\SdTvS \leq 0.01$.
In the present paper electronic spectra and event rates are 
predicted for the \SK\ detector from the probabilities presented in 
\cite{ArticleI}. 
Spectra and event rates are computed for the same set of neutrino parameters 
\dms\ and \SdTvS\ considered in \cite{ArticleI}. 
This short communication reports the main results from these calculations, and 
it is preliminar to a more dettailed paper which presently is under writing 
\cite{ArticleII}.

\section{Formalism}
In \cite{ArticleI}\ neutrino events are classified in four classes or 
``samples'' in accord with their detection time.
The samples are labelled: \DAY, \night, \core\ and \mantle;
where \DAY\ and \night\ samples are composed by those neutrino events 
detected during the day or the night, 
while the other samples are composed by those neutrino events 
which are detected at night and 
which are produced by neutrinos 
which cross the Earth core (\core) 
or does not cross it (\mantle).
Since this classification, all the quantities defined in this paper
(probabilities, spectra, event rates, asymmetries, etc) 
must be referred to one of these four samples.
For this reason each quantity will be denoted by an apex $s = D$, $N$, $C$, $M$\ 
where these letters are for \DAY, \night, \core\ and \mantle\ samples respectivelly.
In this way electronic spectra are denotated by: 
$\SeD(\Te)$, $\SeN(\Te)$, $\SeC(\Te)$, $\SeM(\Te)$\ while event rates are 
denotated by:
\ReD, \ReN, \ReC, \ReM. 
The symbols $\SeZr(\Te)$\ and $\ReZr$\ are for electronic spectra and event rates
computed for massless neutrinos and the Solar Standard Model (SSM).
Few formulae connect these quantities to cross sections, 
neutrino spectra and probabilities. 
For spectra:
\bec\beq\label{eq:electron:spectra:SSM}
{\displaystyle
     \SeZr(\Te) = 
     {\displaystyle \int_{ \Te\left(1 + 
            {{m_e}\over{2 T_e}}\right)}^{+\infty} }
       d\Enu\, \SnuZr(\Enu) \, \dseedEe
},
\eeq\eec

\noindent
and
\bec\beq
\begin{array}{lll}\label{eq:electron:spectra:msw}
    \Ses(\Te) =&  {\displaystyle \int_{ \Te\left(1 +
                     {{m_e}\over{2 T_e}}\right)}^{\displaystyle +\infty}}
                     d\Enu\, 
                     \SnuZr(\Enu) \,
                  &\left[ \dseedEe P_\oplus^s\left({{E_\nu}\over{\Delta m^2}}\right) + \right.\\
         &\\
         && + \left.\dsemdEe 
              \left(1-P_\oplus^s\left({{E_\nu}\over{\Delta m^2}}\right)\right) 
    \right],\\ 
\end{array} 
\eeq\eec

\noindent
with $s = \mbox{$D$, $N$, $C$, $M$}$\ and
where $\Enu$ is the incoming neutrino energy, \Te\ is the emerging electron 
kinetic energy, \elmass\ is the electron mass, $\SnuZr(\Te)$\ is the SSM 
spectrum for \BHt\ neutrinos,
$\PTot^s(\nue \rightarrow \numt)$\ is the transition probability 
for \DAY, \night, \core\ and \mantle.
The cross sections for elastic neutrino scattering over electrons are:
$\dseedEe$\ for \nue\ and  $\dsemdEe$\ for \num, \nut.
The formulae for these cross sections are reported in 
\cite{Bahcall:1989, Bahcall:Kamionkowski:Sirlin:1995}.
Event rates are obtained from:
\bec\beq
\begin{array}{lll}
  \ReZr(\TeTh) &=&  {\displaystyle \int^{+\infty}_{\TeTh} } d\Te\,\SeZr(\Te),\\
   &&\\
  \Res(\TeTh) &= & {\displaystyle \int^{+\infty}_{\TeTh} } d\Te \, \Ses(\Te).\\
\end{array}
\eeq\eec

\noindent
Here it is assumed
that event rates are computed for an electron kinetic energy threshold \TeTh\
of 5 MeV, while event rates are expressed in units of the SSM prediction so
that $\ReZr \equiv 1$.

Spectra and event rates may be combined to produce various indicators usefull
for the experimental data analysis. Three indicators are considered here: 
spectral distortion, spectral D - N asymmetry and event rates \daynight\ 
asymmetry.
The spectral distortion compares the predicted electronic spectra 
for the MSW effect (or the measured one) with the spectra predicted for 
the SSM and the massless neutrino. It is defined as:
\bec\beq
\delta \Ses(\Te) = \frac{\Ses(\Te)}{\SeZr(\Te)}, \mbox{\hspace{1cm}} 
            s = \mbox{$D$, $N$, $C$, $M$}.
\eeq\eec

\noindent
In absence of any systematical effect this index has a simple interpretation: 
if $\delta \Ses(\Te) \equiv 1$\ there is no MSW effect and the SSM is correct, 
if $\delta \Ses(\Te)$\ is a constant but it is different from 1 then there 
is an energy independent neutrino effect or the SSM is wrongth.
At last, if $\delta \Ses$\ changes with the energy then an energy dependent 
neutrino effect occurs, which should be a sign of a new physics.
The spectral \daynight\ asymmetry is an indicator of the \daynight\ effect.
It is defined as:
\bec\beq
     \AsymSs(\Te) = 2 \frac{\Ses(\Te)-\SeD(\Te)}{\Ses(\Te)+\SeD(\Te)},
      \mbox{\hspace{1cm}}
       s = \mbox{$N$, $C$, $M$},
\eeq\eec

\noindent
a zero asymmetry is an indication of no-Earth effect.
Finally, the Event Rate \daynight\ asymmetry (or shortly the {\em asymmetry}) is 
defined as:
\bec\beq
     \calA_R^s = 2 \frac{\Res - \ReD}{\Res+\ReD}, 
      \mbox{\hspace{1cm}}
       s= \mbox{$N$, $C$, $M$}.
\eeq\eec

\section{Results}
Figure 1, Figures from 2.1 to 2.18 and table I shows the predictions 
for spectra, event rates and related distortions and asymmetries 
for the selected set of neutrino parameters \dms\ and \SdTvS.
Since the MSW effect in the Earth is not large, its influence on the 
electronic spectra are better illustrated by spectral distortions and 
asymmetries than by the electronic spectra itself.
So, in the present paper, only one set of spectra is reported in Fig. 1 for 
$\SdTvS = 0.01$\ and $\dms = 5 \times 10^{-5}$\ eV$^2$.
In the figure the upper full - line is the SSM spectra while 
$\SeD(\Te)$, $\SeN(\Te)$, $\SeC(\Te)$\ and $\SeM(\Te)$\ are represented 
respectively by the long - dashed line, the short - dashed line, the lower full
line and the dotted line.

Figures from 2.1 to 2.18 reports 
the spectral distortions (upper frame of each figure)
and the spectral D-N asymmetries (lower frame of each figure) 
predicted for each combination of the MSW parameters \dms\ and \SdTvS\ listed in
Tab. I.
The enhancement in the spectral distortions and in the \daynight\
asymmetries introduced by the \core\ sample is fairly evident,
expecially for $\SdTvS \lsim 0.01$.
The magnitude of the effect is sensitive to \dms.
Figures from 2.1 to 2.18 depict also many other features like: kinks, knees
and 
peaks which at last are associated to equivalent features displayed in the 
probabilities reported in \cite{ArticleI}.

The predicted 
event rates associated to electronic 
spectra in Figures from 2.1 to  2.18 are listed in 
Tab. I.
Each electronic spectra is associated with one entry in the table through the
numbers in the first column.
Event rates expressed in units of \ReZr\ are listed in columns from 4$^{th}$\ to 
7$^{th}$.
Percentual event rates asymmetries ($\calA_R^s \times 100$)\ are listed in 
columns from 8$^{th}$\ to 10$^{th}$. 
It is evident that even for the event rates the predicted asymmetries are
enhanced
when the \core\ sample is extracted by the full set of night neutrinos. 
This is well displayed by the last column which shows the ratio:
$|\AsymRC| / |\AsymRN|$. 
The enhancement may be as high as a factor of order six.
This enhancement should increase the 
sensitivity of the \SK\ detector to the 
\daynight\ effect, in the case the small mixing angle 
solution to the solar neutrino problem is correct.
From the table it is evident that it is theorethically possible to have a 
significative asymmetry, at least till $\SdTvS \approx 0.006$ and 
at least for $5 \times 10^{-6}$ eV$^2 \,\lsim\, \dms \,\lsim\, 10^{-5}$ eV$^2$.
As expected from probabilities in \cite{ArticleI}, a negative Earth effect 
for $\SdTvS < 0.004$\ is also present. 

\section{Conclusions}
In a previous work \cite{ArticleI}\
 a set of probabilities for the MSW transition 
$\nue \rightarrow \numt$\ were computed for solar neutrinos traversing the
Sun and the Earth.
From them, a set of electronic spectra,
event rates and related \daynight\ 
asymmetries is predicted for the \SK\ detector. The computation was 
extended over a comphensive set of neutrino parameters \dms\ and \SdTvS.

From these predictions it comes that the \SK\ detector should be sensitive 
to the \daynight\ effect even for $\SdTvS < 0.01$
and $5 \times 10^{-6}$ eV$^2 \,\lsim\, \dms \,\lsim\, 10^{-5}$ eV$^2$.
As expected also by previous studies 
\cite{Baltz:Weneser:1994, Krastev:1996, Gelb:Kwong:Rosen:1996},
this sensitivity may be reached selecting those neutrino events 
which are induced 
in the \SK\ detector by neutrinos whose trajectories cross the Earth core.
Our results suggests an enhancement up to a factor six 
in the \daynight\ asymmetry for the \SK\ detector.

At last we recall that this is a short communication only, which precedes
a more complete and dettailled paper presently under writing \cite{ArticleII}.

\section{Acknowledgments}
M.M. wishes to thank the International School for Advanced Studies, Trieste,
Italy, where part of the work for this study has been done, for kind
hospitality and financial support.
The authors are indebted to the ICARUS group of the University of Pavia  
and INFN, Sezione di Pavia, and especially to Prof. E. Calligarich, 
for allowing the use of their computing facilities for the present study.
M.M. wishes to thank also Dr. A. Rappoldi for his suggestions
concernig the computational aspects of the study, and to Prof. A. Piazzoli
for his constant interest in the work and support.



\newpage




\bec           
\begin{tabular}{|rll|llll|rrr|r|}
\multicolumn{11}{c}{{\bf{Tab. \TabEventRates:} Event Rates and
 D - N Asymmetries for \SK}}\\
\hline
\multicolumn{3}{|l}{}&\multicolumn{4}{|c}{Event Rates}&
                               \multicolumn{3}{|c|}{$\AsymR \times 100$}&\\
\multicolumn{1}{|c}{N.}& \multicolumn{1}{c}{$\SdTvS$}&
\multicolumn{1}{c|}{$\dms$}      &
\multicolumn{1}{c}{\DAY} &
\multicolumn{1}{c}{\night}      &
\multicolumn{1}{c}{\core}      &
\multicolumn{1}{c}{\mantle}    &
\multicolumn{1}{|c}{\night}      &
\multicolumn{1}{c}{\core}      &
\multicolumn{1}{c}{\mantle}    &
\multicolumn{1}{|c|}{$\frac{|\AsymRC|}{|\AsymRN|}$}    \\
    \hline\hline
1 & 0.0008 & 9.0e-6  & 0.8441 & 0.8431 & 0.8379 & 0.8440 & -0.12 & -0.75 & -0.01 & 6.25\\
2 & 0.0008 & 7.0e-6  & 0.8742 & 0.8722 & 0.8625 & 0.8739 & -0.22 & -1.35 & -0.04 & 6.14\\
3 & 0.0008 & 5.0e-6  & 0.9052 & 0.9020 & 0.8949 & 0.9031 & -0.35 & -1.14 & -0.23 & 3.26\\
    \hline
4 & 0.0010 & 9.0e-5  & 0.2413 & 0.2413 & 0.2413 & 0.2413 & 3e-3  & 4e-3  & 3e-3  & 1.33\\
5 & 0.0010 & 7.0e-6  & 0.8472 & 0.8450 & 0.8338 & 0.8468 & -0.26 & -1.59 & -0.04 & 6.12\\
6 & 0.0010 & 5.0e-6  & 0.8846 & 0.8809 & 0.8727 & 0.8822 & -0.43 & -1.35 & -0.27 & 3.14\\
    \hline
7 & 0.0020 & 1.0e-5  & 0.6426 & 0.6419 & 0.6386 & 0.6425 & -0.10 & -0.62 & -0.01 & 6.20\\
8 & 0.0020 & 7.0e-6  & 0.7270 & 0.7244 & 0.7113 & 0.7266 & -0.36 & -2.18 & -0.07 & 6.06\\
9 & 0.0020 & 5.0e-6  & 0.7905 & 0.7852 & 0.7747 & 0.7869 & -0.67 & -2.02 & -0.45 & 3.01\\
    \hline
10 & 0.0040 & 1.0e-5  & 0.4408 & 0.4417 & 0.4465 & 0.4410  & 0.22  & 1.28 &  0.04 & 5.82\\
11 & 0.0040 & 7.0e-6  & 0.5469 & 0.5470 & 0.5481 & 0.5468  & 0.01  & 0.21 & -0.02 & 21.0\\ 
12 & 0.0040 & 5.0e-6  & 0.6378 & 0.6343 & 0.6307 & 0.6349  & -0.56 & -1.12 & -0.47 & 2.00\\
    \hline
13 & 0.0060 & 1.0e-5  & 0.3233 & 0.3267 & 0.3430 & 0.3239 & 1.05 & 6.20 & 0.17 & 5.90\\ 
14 & 0.0060 & 7.0e-6  & 0.4243 & 0.4297 & 0.4578 & 0.4251 & 1.26 & 7.60 & 0.17 & 6.03\\
15 & 0.0060 & 5.0e-6  & 0.5224 & 0.5248 & 0.5374 & 0.5228 & 0.46 & 2.82 & 0.06 & 6.13\\
    \hline
16 & 0.0080 & 1.0e-5  & 0.2543 & 0.2603 & 0.2913 & 0.2552 & 2.36 & 13.59 & 0.37 & 5.76\\ 
17 & 0.0080 & 7.0e-6  & 0.3405 & 0.3523 & 0.4125 & 0.3423 & 3.39 & 19.10 & 0.53 & 5.63\\ 
18 & 0.0080 & 5.0e-6  & 0.4351 & 0.4458 & 0.4787 & 0.4404 & 2.44 & 9.54  & 1.22 & 3.91\\
    \hline
19 & 0.0100 & 7.0e-6  & 0.2830 & 0.3013 & 0.3942 & 0.2860 & 6.28 & 32.85 & 1.05 & 5.23\\
20 & 0.0100 & 5.0e-6  & 0.3688 & 0.3891 & 0.4435 & 0.3802 & 5.36 & 18.38 & 3.03 & 3.43\\
    \hline
21 & 0.0130 & 5.0e-6  & 0.2979 & 0.3333 & 0.4188 & 0.3192 & 11.21 & 33.72 & 6.89 & 3.01\\
    \hline
    \hline
22 & 0.3000 & 1.5e-5  & 0.2171 & 0.2424 & 0.2530 & 0.2407 & 11.03 & 15.29 & 10.30 & 1.39\\
23 & 0.3000 & 2.0e-5    & 0.2181 & 0.2357 & 0.2426 & 0.2346 &  7.79 & 10.66 & 7.31 & 1.37\\
24 & 0.3000 & 3.0e-5    & 0.2214 & 0.2323 & 0.2348 & 0.2319 &  4.78 & 5.85  & 4.61 & 1.22\\
25 & 0.3000 & 4.0e-5    & 0.2275 & 0.2352 & 0.2370 & 0.2349 &  3.31 & 4.07 & 3.19  & 1.23\\
    \hline
26 & 0.4800 & 3.0e-5   & 0.2717 & 0.2885 & 0.2916 & 0.2879 & 5.99 & 7.08 & 5.81 & 1.18\\
27 & 0.4800 & 5.0e-5   & 0.2887 & 0.2975 & 0.2992 & 0.2972 & 2.99 & 3.57 & 2.89 & 1.19\\
    \hline
28 & 0.5000  & 2.0e-5   & 0.2735 & 0.3011 & 0.3078 & 0.3000 & 9.60 & 11.79 & 9.23 & 1.23\\
    \hline
29 & 0.5600 & 1.0e-5   & 0.2800 & 0.3693 & 0.4234 & 0.3604 & 27.50  & 40.77 & 25.11 & 1.48\\
    \hline
30 & 0.6000  & 8.0e-5   & 0.3685 & 0.3738 & 0.3746 & 0.3736 & 1.42 & 1.65 & 1.38 & 1.16\\
    \hline
31 & 0.7000  & 3.0e-5   & 0.3449 & 0.3683 & 0.3715 & 0.3677 & 6.56 & 7.43 & 6.41 & 1.13\\
32 & 0.7000  & 5.0e-5   & 0.3588 & 0.3712 & 0.3733 & 0.3708 & 3.38 & 3.95 & 3.29 & 1.17\\
    \hline
33 & 0.7700 & 2.0e-5   & 0.3697 & 0.4083 & 0.4110 & 0.4079 & 9.94 & 10.59 & 9.83 & 1.07\\
    \hline
34 & 0.8000  & 1.3e-4 & 0.4744 & 0.4771 & 0.4777 & 0.4770 & 0.57 & 0.69 & 0.55 & 1.21\\
   \hline
\end{tabular}
\eec

\newpage

\bec{\Large\bf{Figure Captions}}\eec


\vspace{1cm}

\noindent
{\bf{Figure \FigSKSpectra:}}
Electronic spectra for \SK.
Spectra are computed for $\SdTvS = 0.01$\ and 
$\dms = 5 \times 10^{-5}$\ eV$^2$, in the plot spectra are normalized 
in units of the SSM event rate for \SK\ per MeV$^{-1}$.
The {\em Upper Full} line is the \standard\ spectrum $\SeZr(\Te)$\
while the other lines are for spectra whith matter effect.
The {\em Long Dashed} line denotes the \DAY\ spectrum,
the {\em Short Dashed} line the \night\ spectrum,
the {\em Full} line is for the \core, while {\em Dots} are for \mantle\
sample.

\vspace{1cm}

\noindent
{\bf{Figure \FigSKDist:}}
Electronic spectral distortion and asymmetries for \SK. 
In the upper frame of each figure it is draft the spectra deformation
in the lower the related asymmetry as a function of \Te.
The number on the top left of each figure associates it with a row of table
\TabEventRates. 
         {\em Long Dashed} lines mark the \DAY\ spectrum distotion
(D-N asymmetry)
         {\em Short Dashed} lines the \night\ spectrum distotion 
(D-N          asymmetry),
         {\em Full} lines are for \core\, while {\em Dotts} are for \mantle.

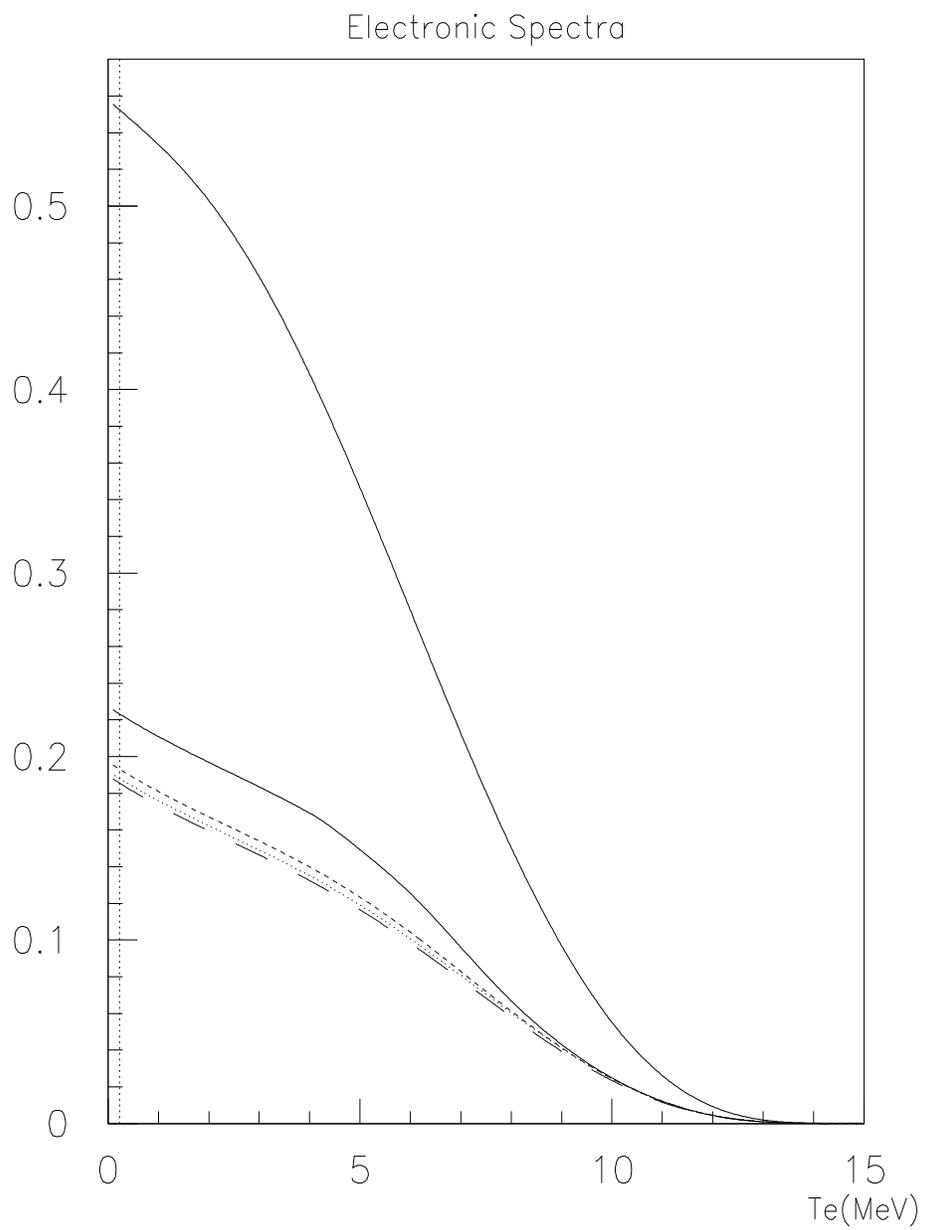
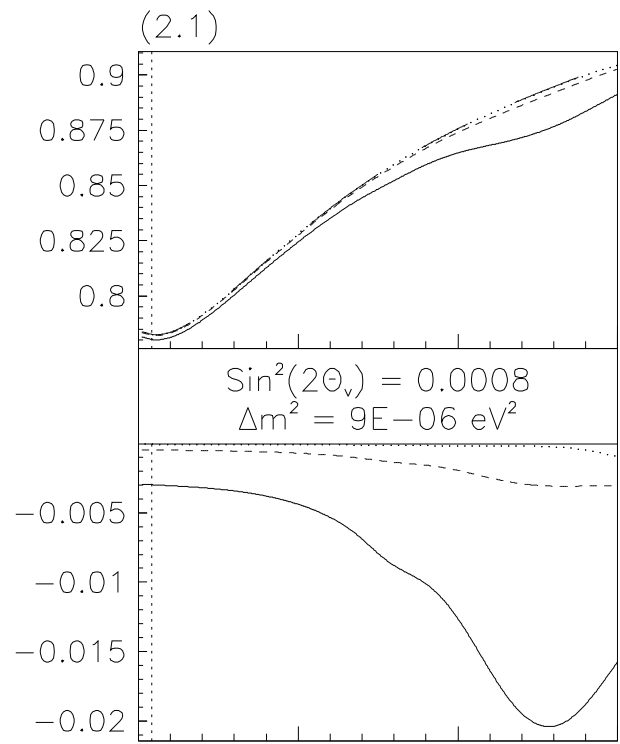
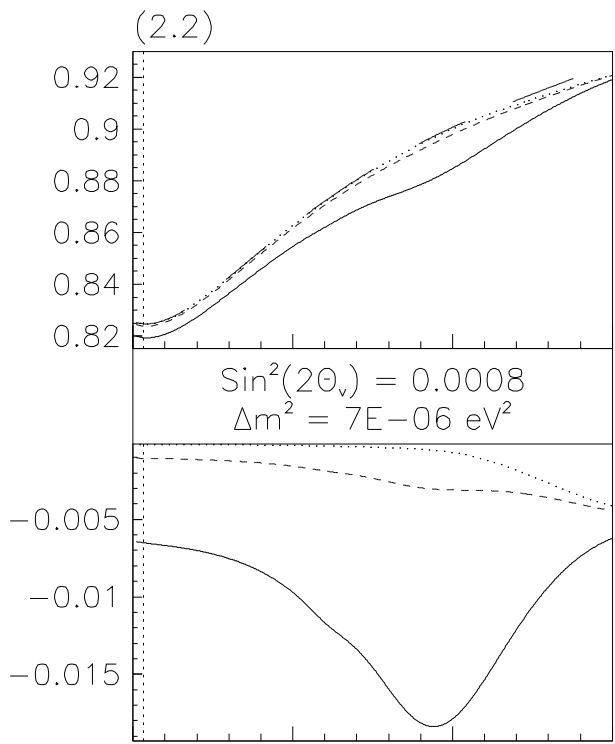
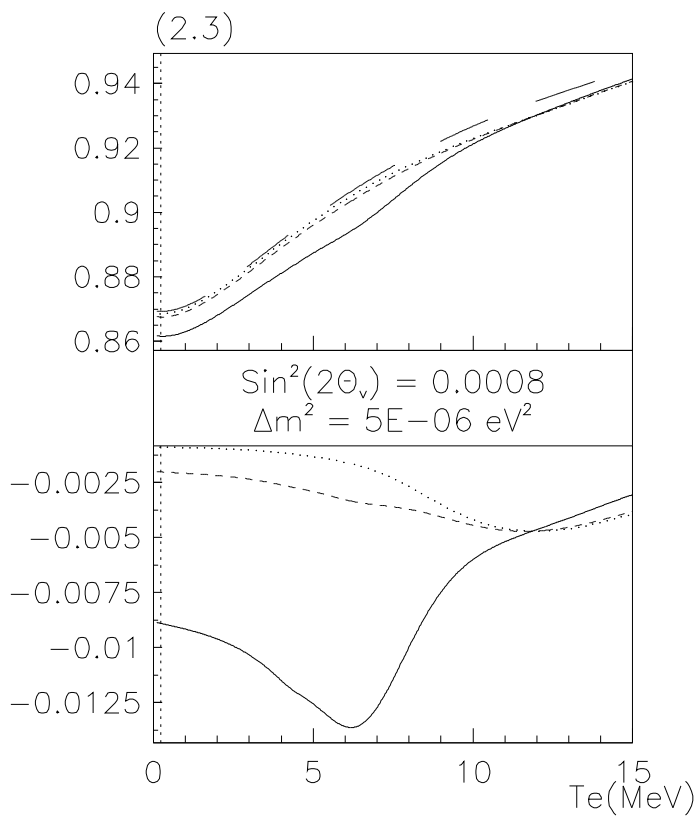
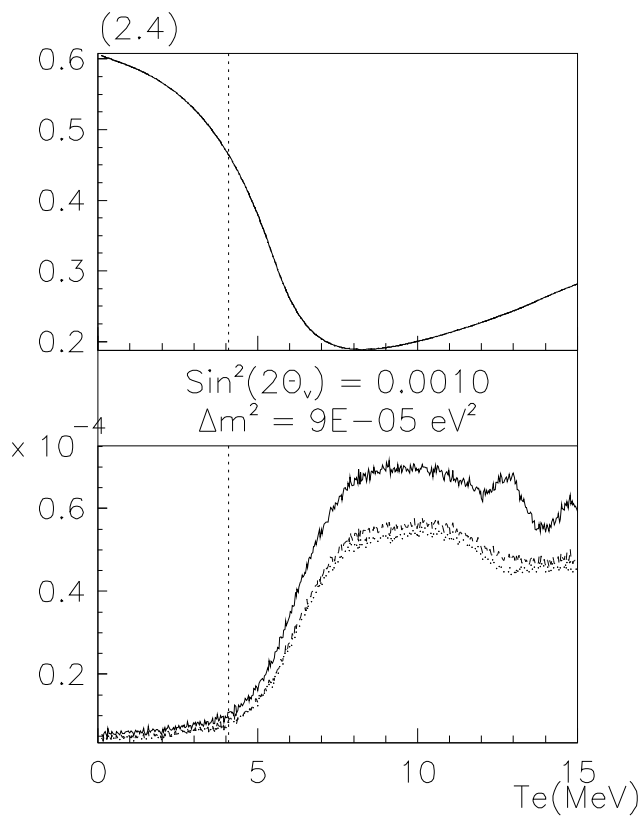
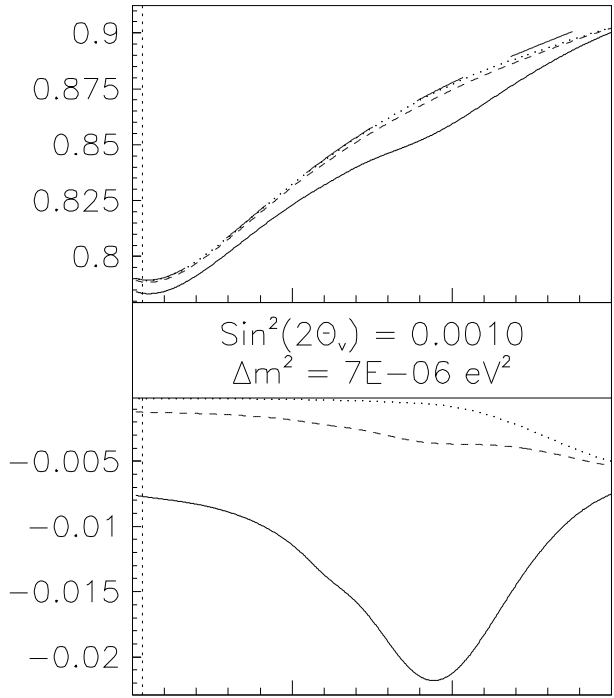
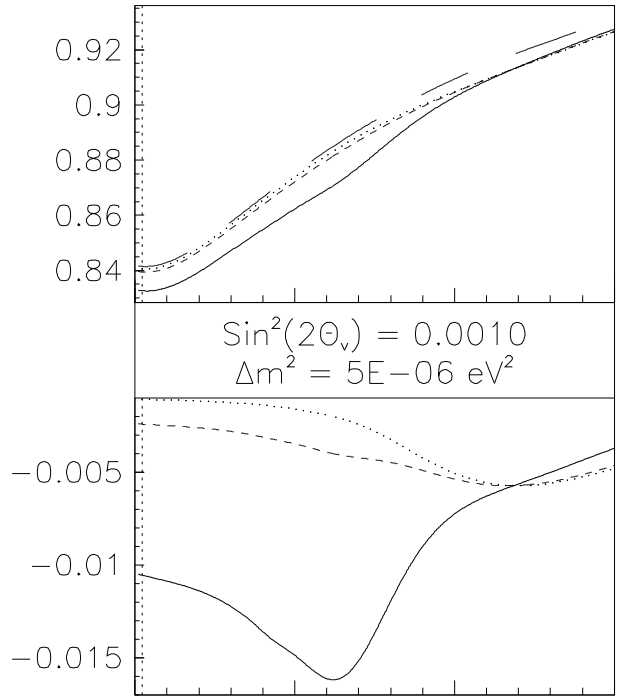
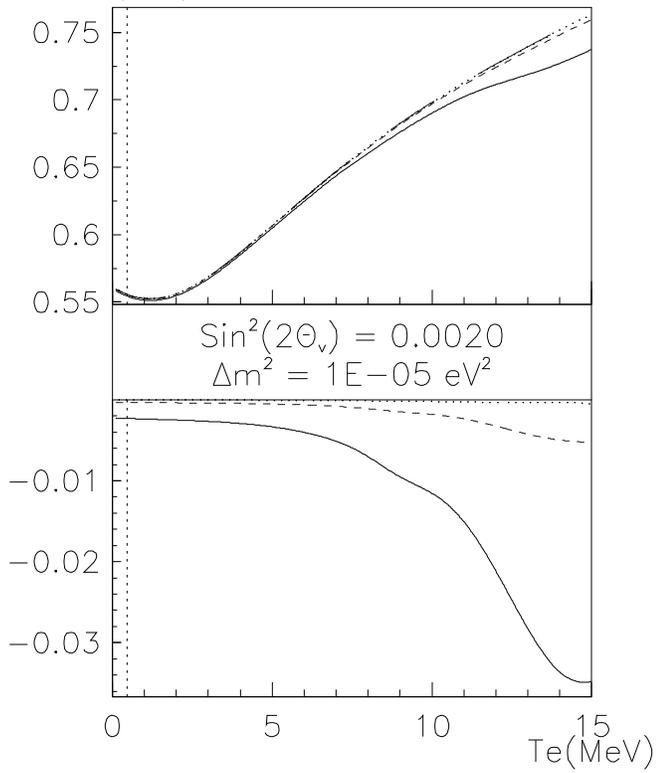
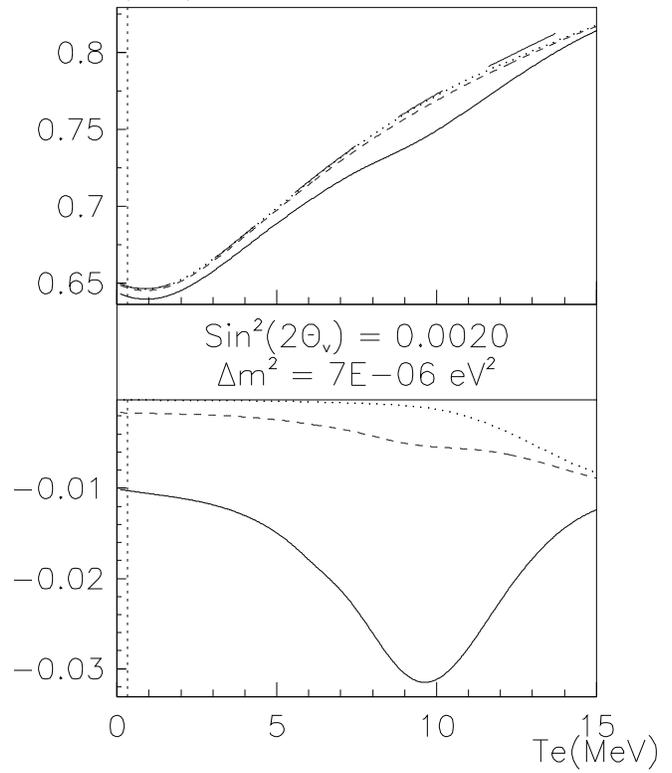
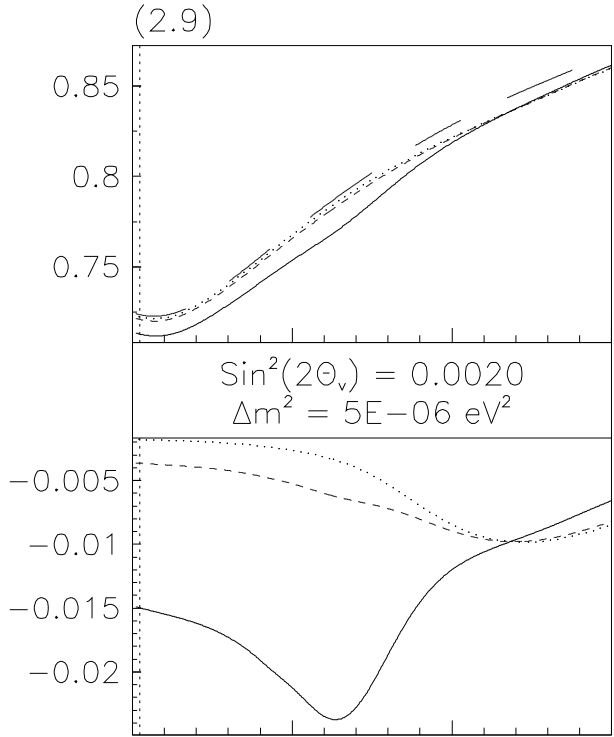
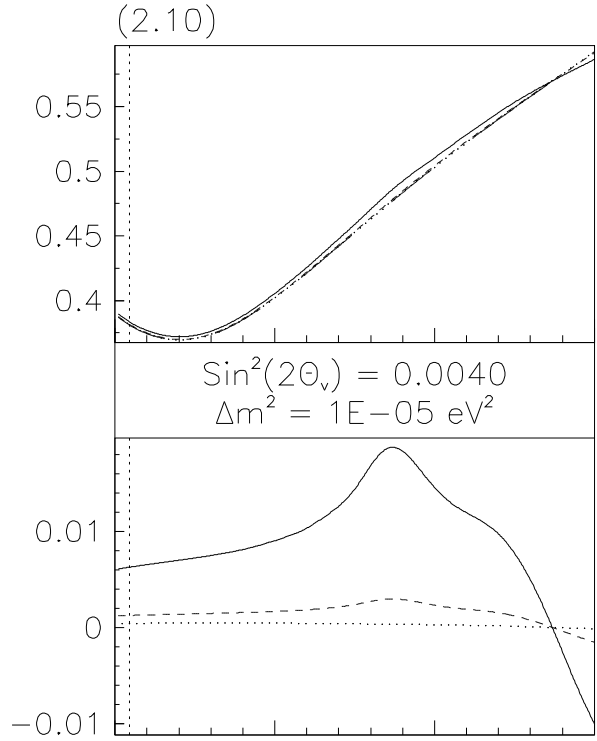
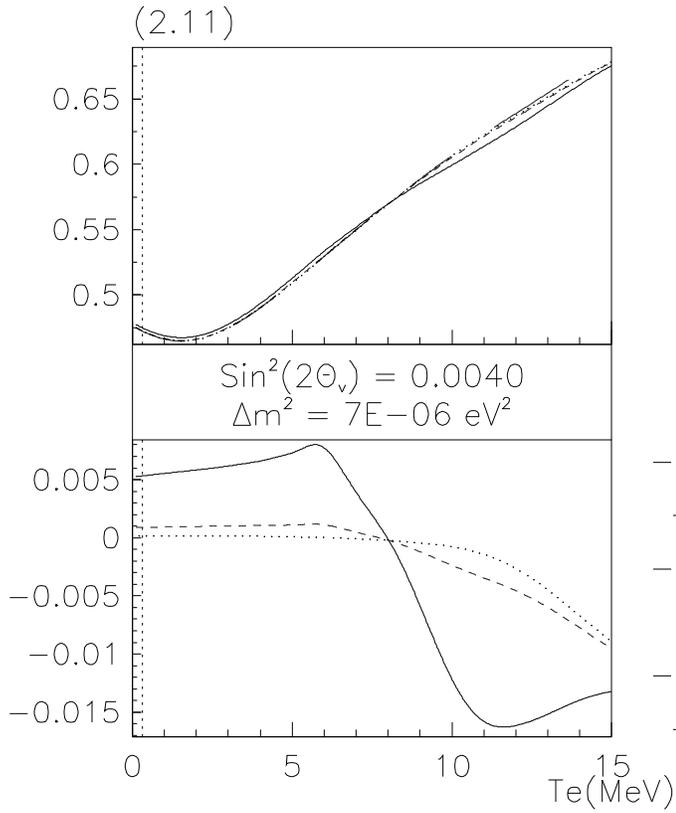
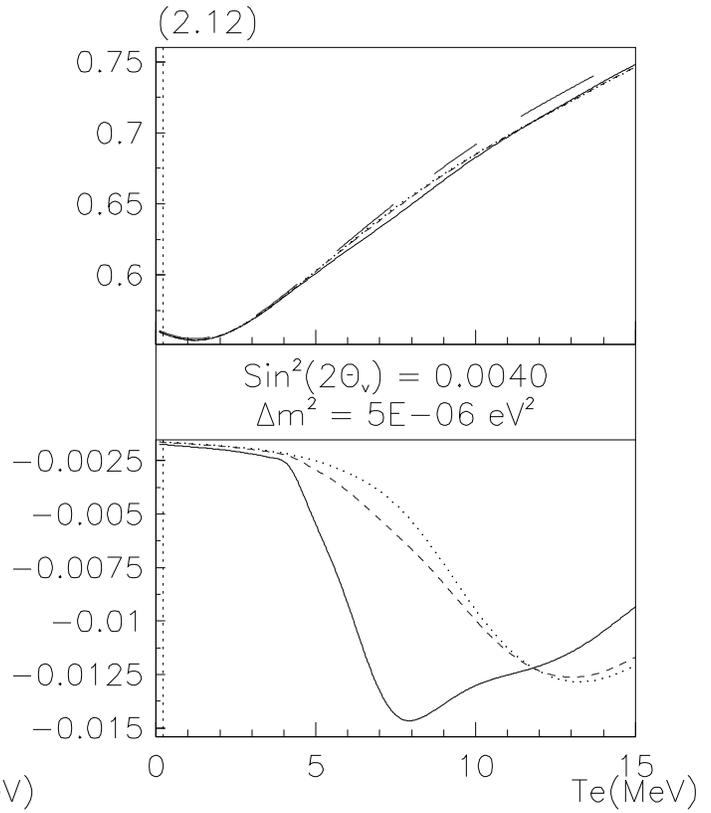
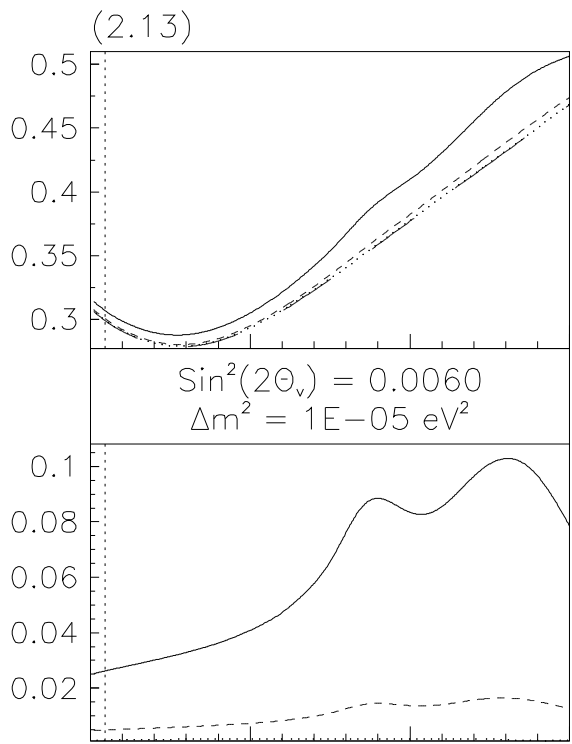
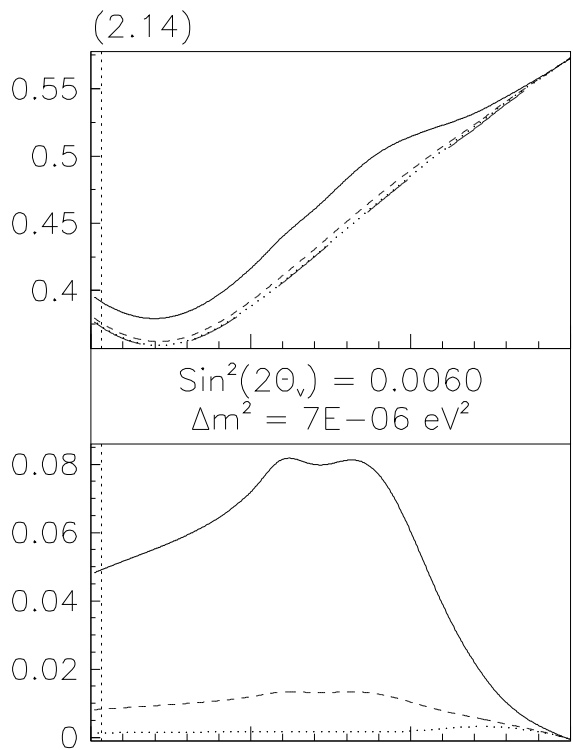
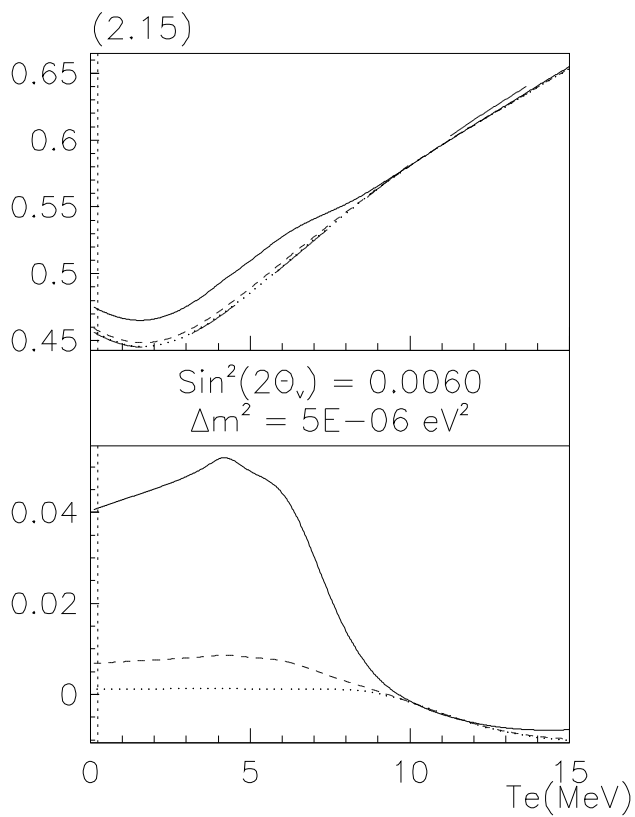
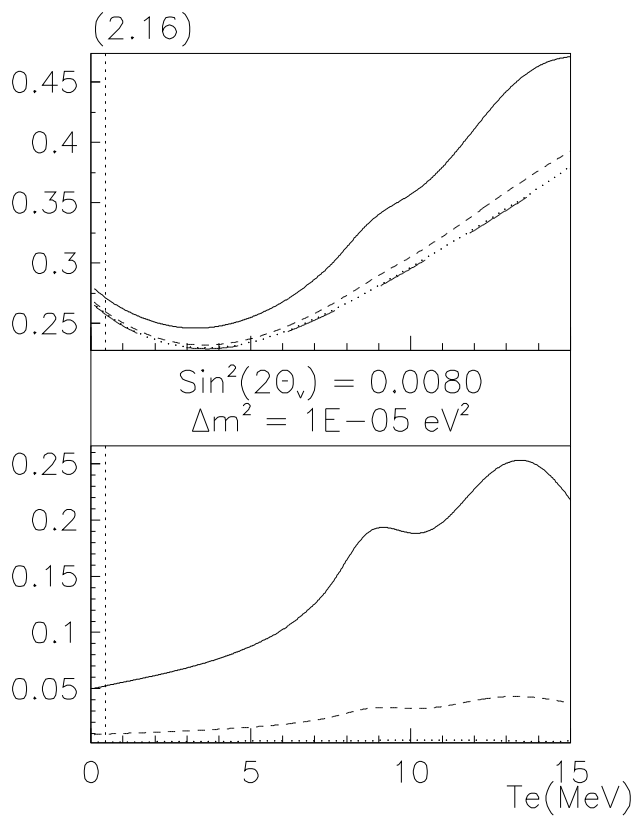
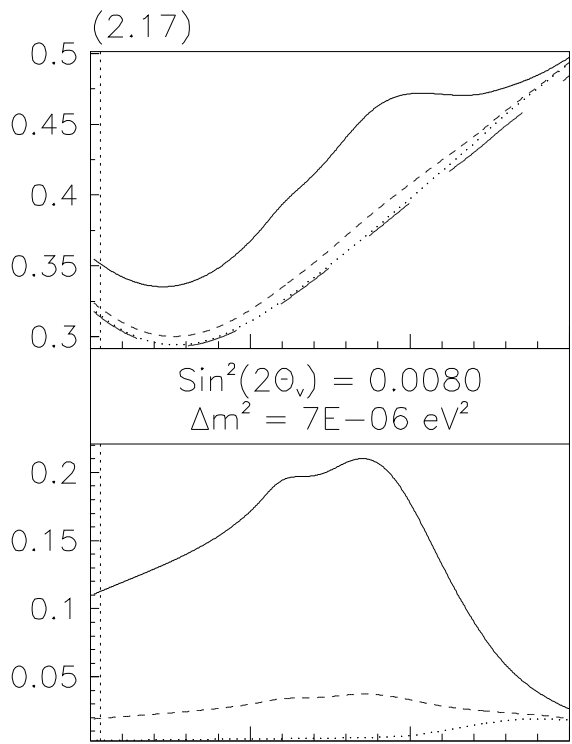
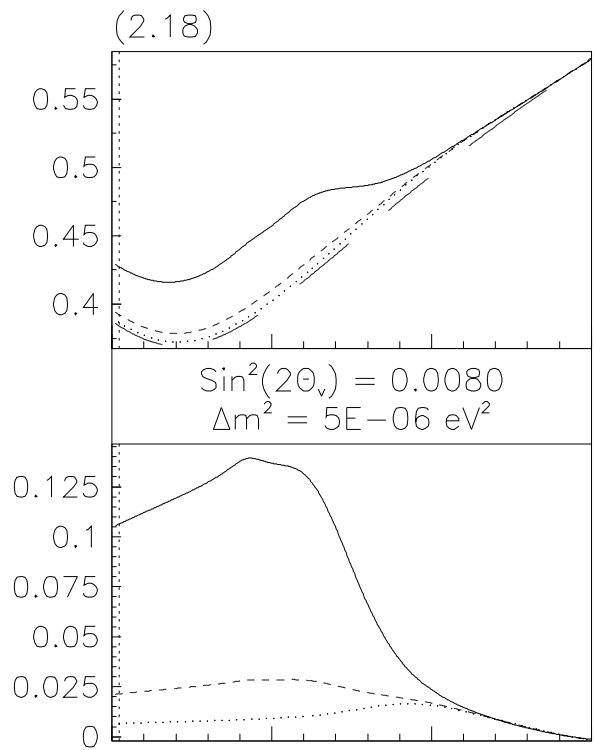
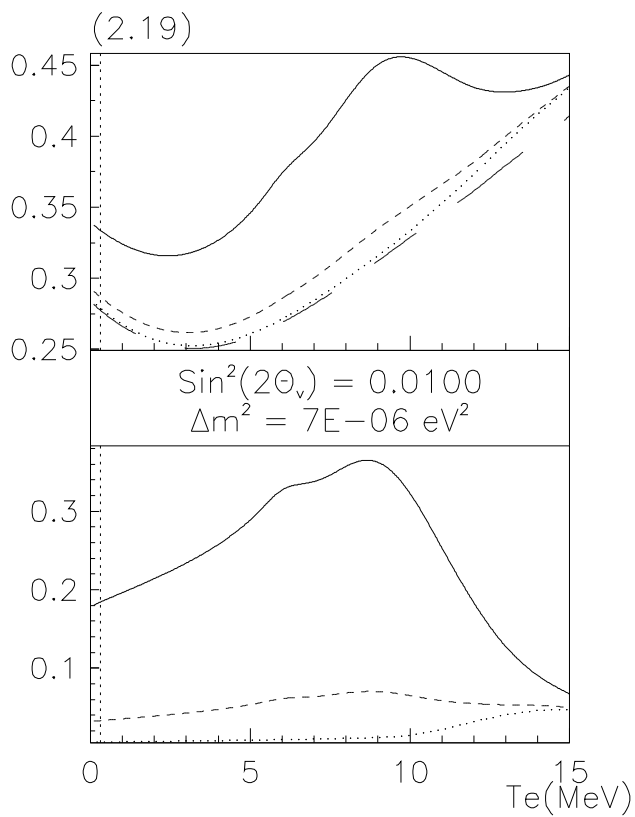
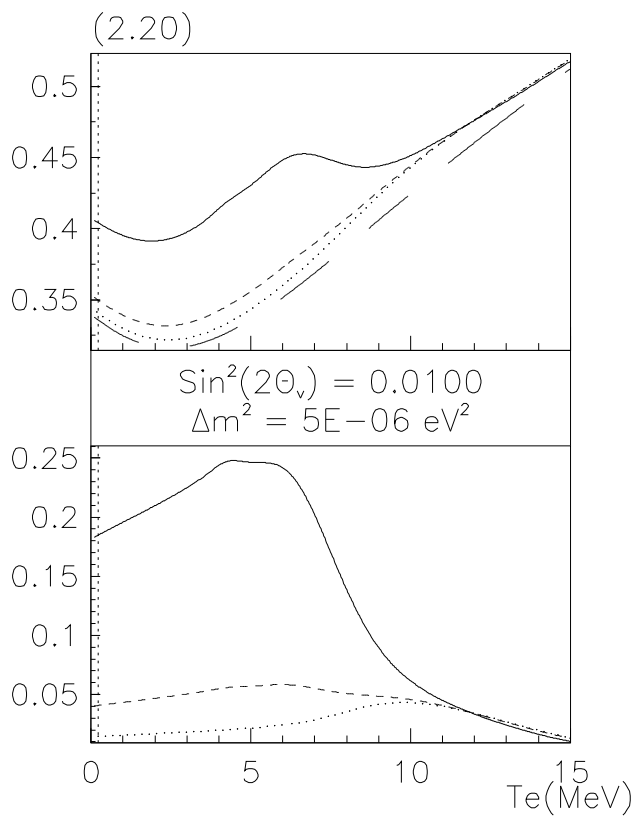
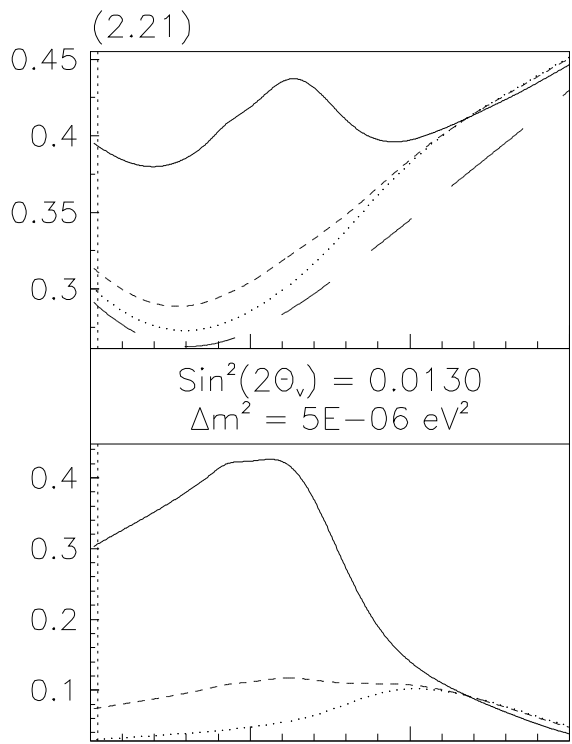
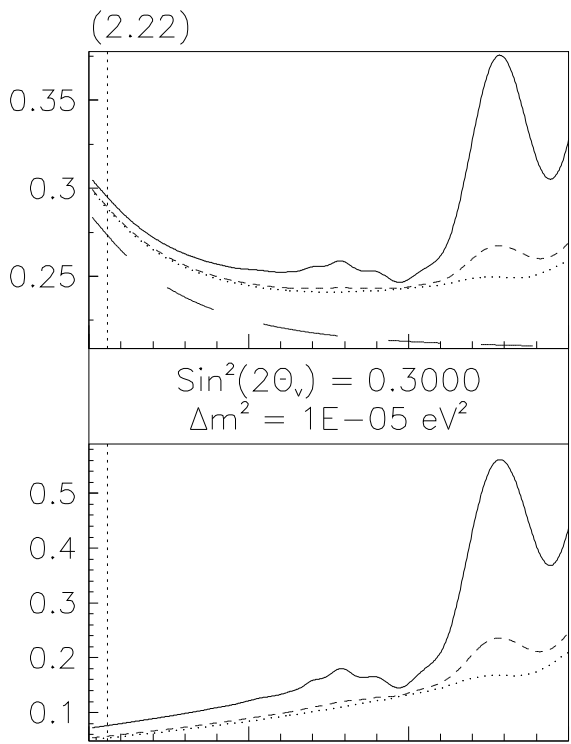
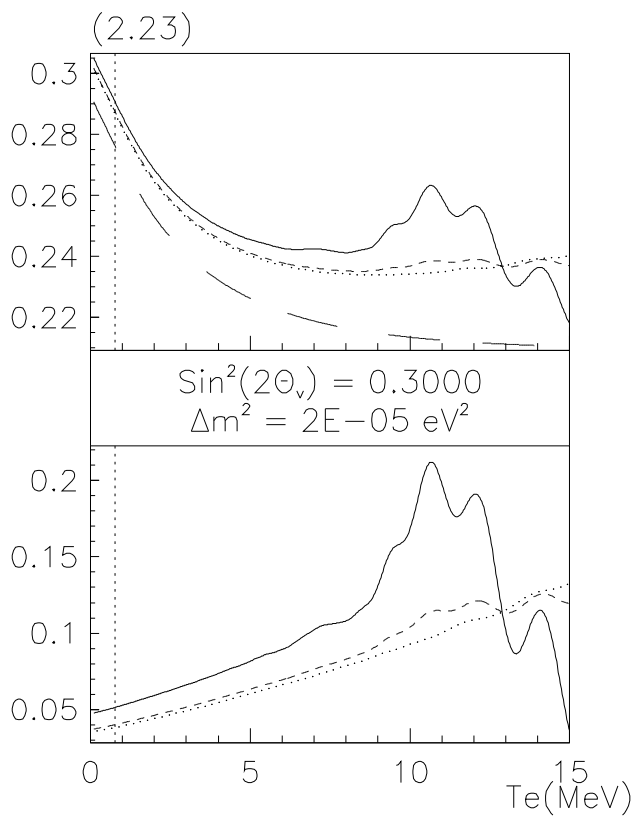
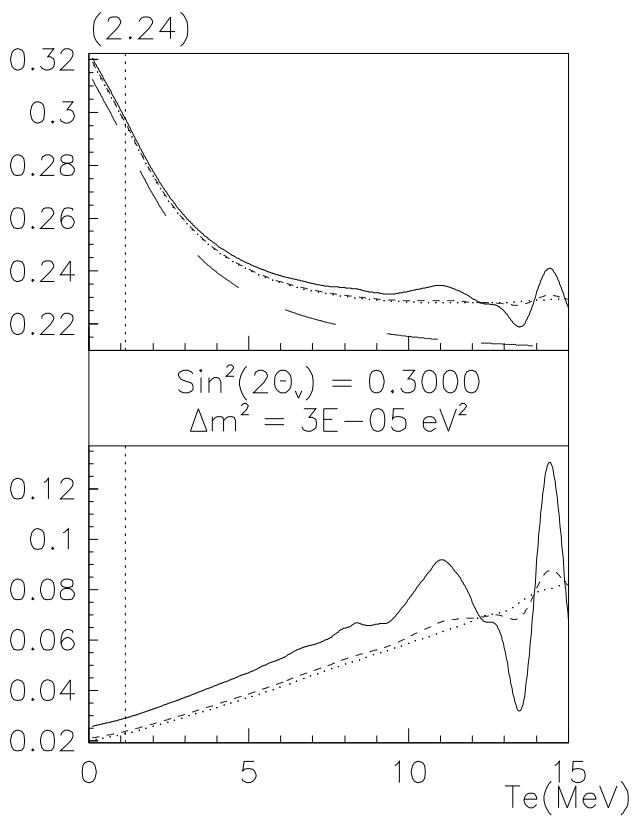
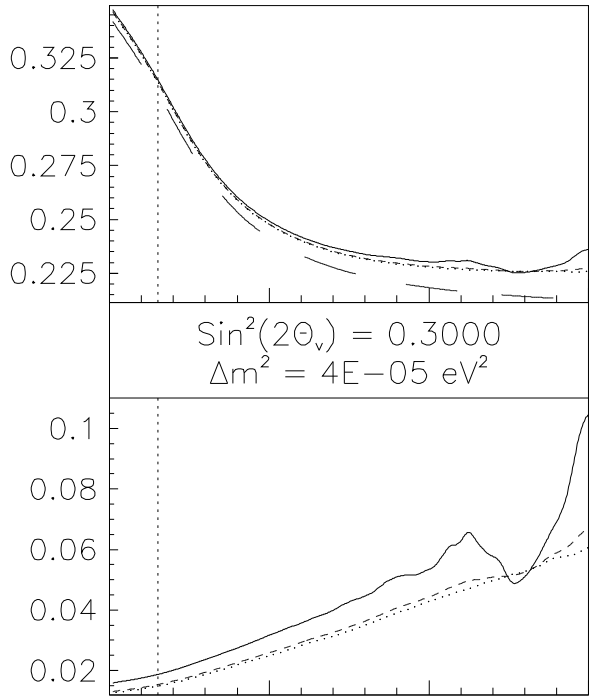
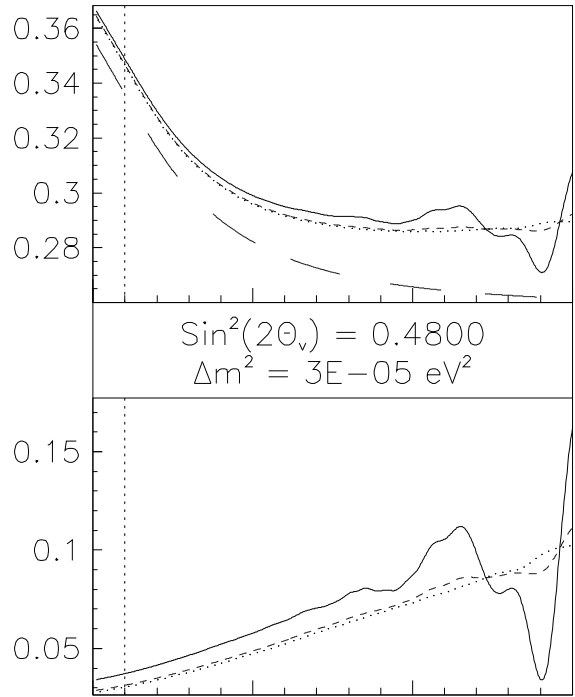
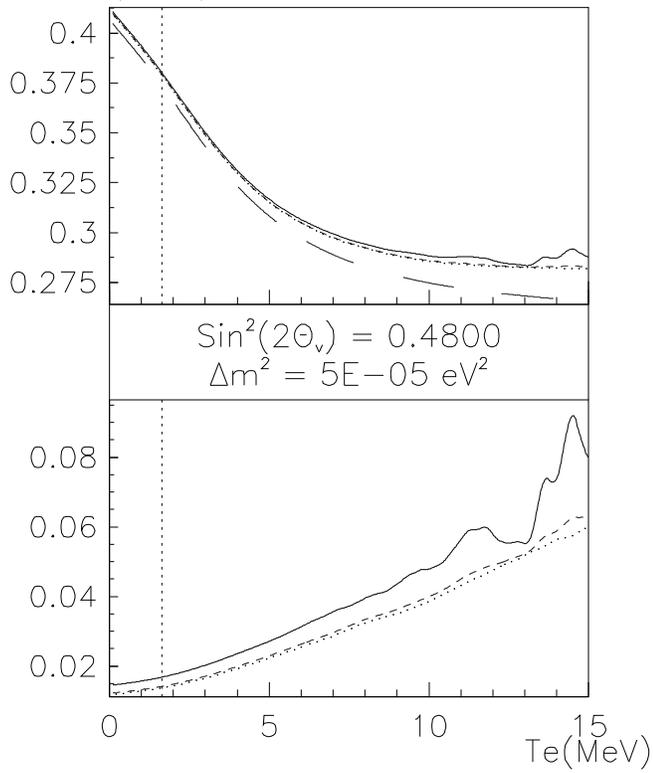
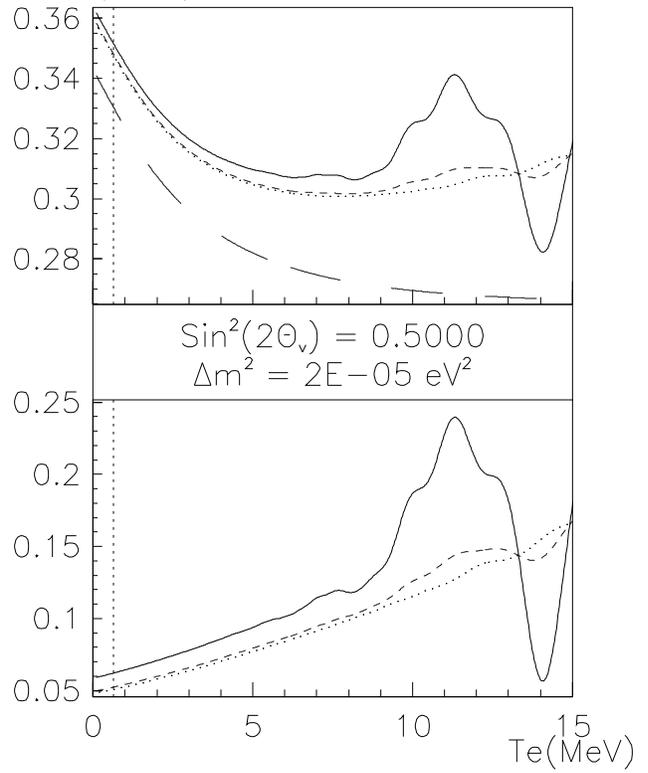
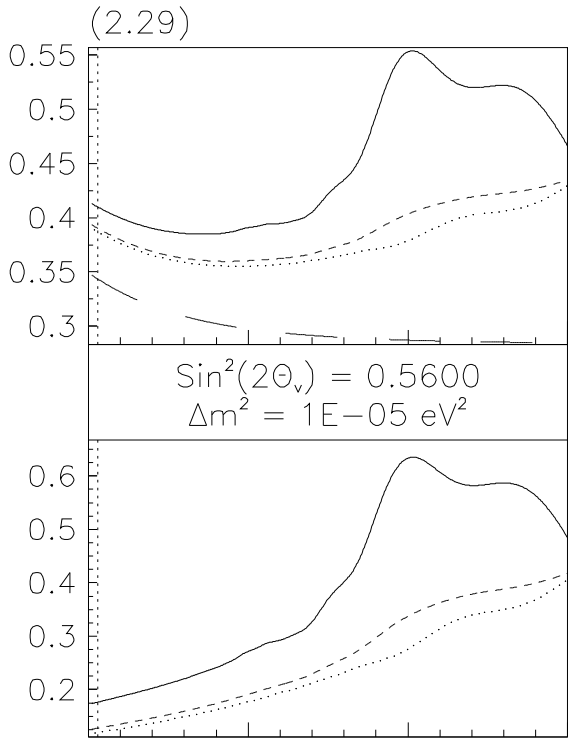
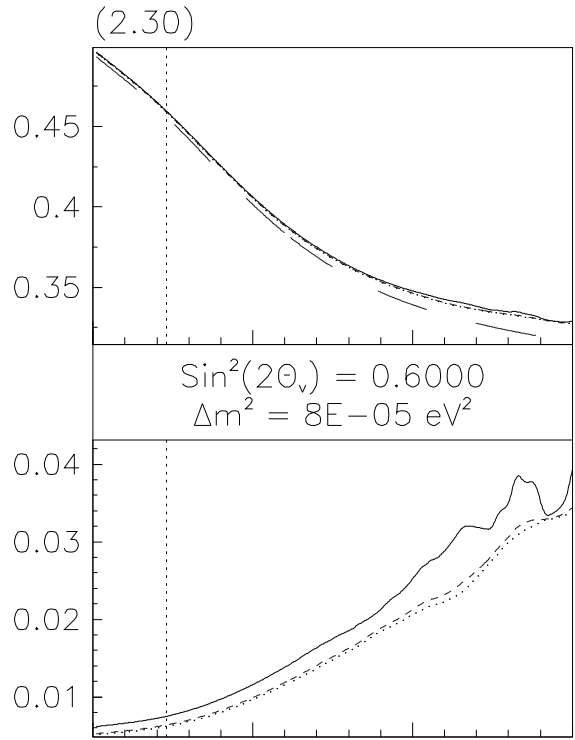
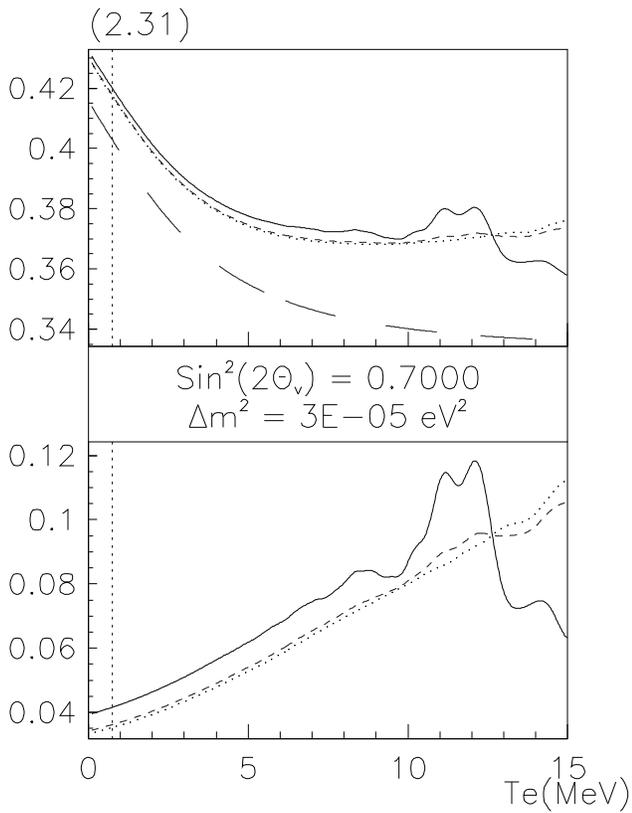
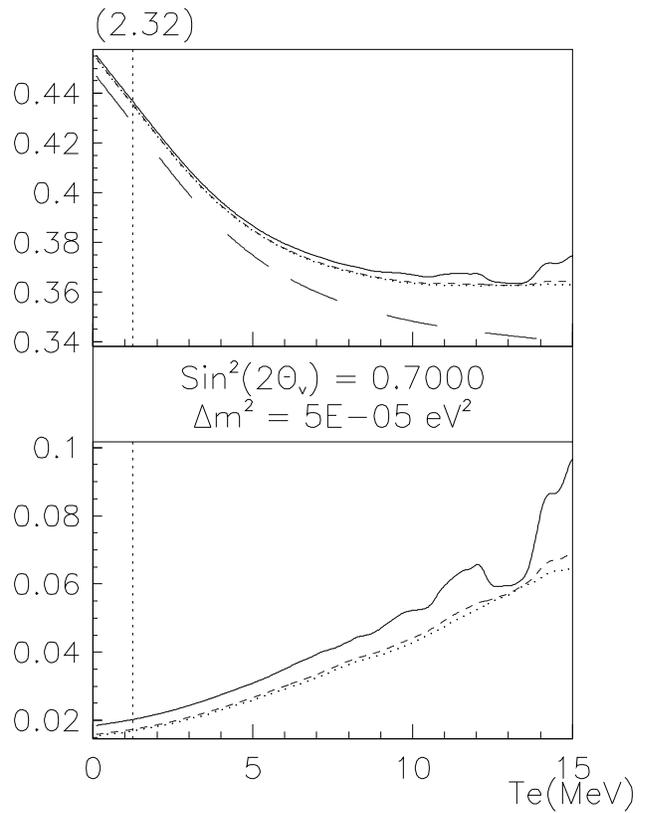
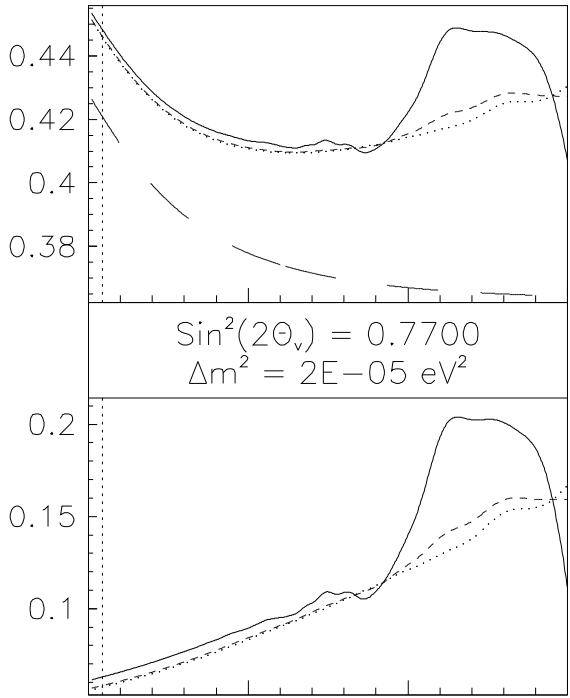
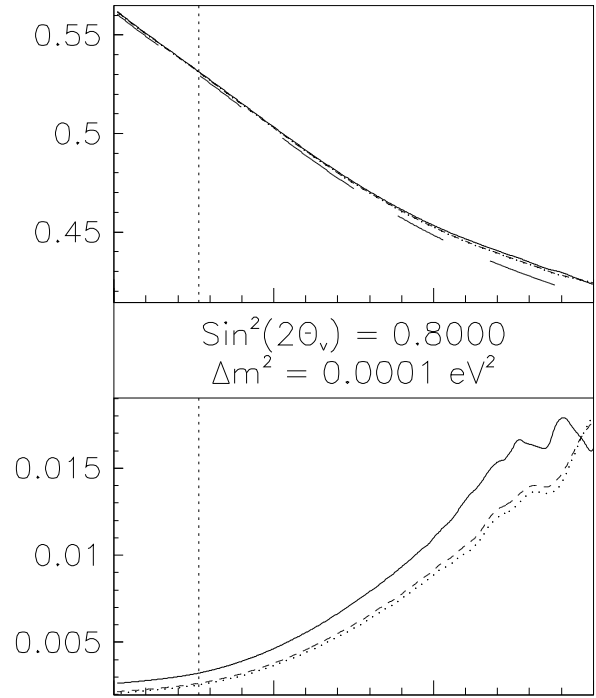
\end{document}